\documentclass{PoS}

\usepackage{lineno}
\usepackage{soul}

\title{$\bf{K^{*}(892)^0}$ production in p+p interactions at 158 GeV/c from NA61/SHINE}

\ShortTitle{$K^{*}(892)^0$ production in p+p interactions at 158 GeV/c from NA61/SHINE}

\author{\speaker{Angelika Tefelska}\thanks{This work was supported by the National Science Centre, Poland (grant 2017/25/N/ST2/02575) and partially supported by the National Science Centre, Poland (grant 2015/18/M/ST2/00125) and the
		Ministry of Science and Higher Education, Poland (DIR/WK/2016/2017/10-1).}  \hspace{0.1cm} for the  NA61/SHINE Collaboration \\
        Warsaw University of Technology, Faculty of Physics\\
        E-mail: \email{angelika.tefelska@cern.ch}}

\abstract{
The measurement of $K^{*}(892)^0$ resonance production via its $K^{+}\pi^{-}$ decay mode in inelastic p+p collisions at beam momentum 158~GeV/c ($\sqrt{s_{NN}}=17.3$~GeV) is presented. The data were recorded by the NA61/SHINE hadron spectrometer at the CERN Super Proton Synchrotron. The first ever double differential measurements and $p_T$-integrated spectra of $K^{*}(892)^0$ at beam momenta of 158 GeV/c was done by using the \textit{template} fitting method. The full phase-space yields, mass and 
width of $K^{*}(892)^0$ mesons are compared with Hadron Resonance Gas models as well as with world data on p+p and nucleus-nucleus collisions.}

\FullConference{Corfu Summer Institute 2018 "School and Workshops on Elementary Particle Physics and Gravity"\\
	(CORFU2018)\\
	31 August - 28 September, 2018\\
	Corfu, Greece}

\begin{document}
\section{Introduction}

The study of short-lifetime resonances are unique tools to understand the less known aspects
of high energy collisions, especially its time evolution. The measurement of $K^{*}(892)^0$ meson production may help to distinguish between two possible scenarios for the fireball freeze-out: the sudden and the gradual one \cite{markert}. The ratio of $K^{*}(892)^0$ to charged kaon production may allow to determine the time between chemical (vanishing inelastic collisions) and kinetic (vanishing elastic collisions) freeze-out. The lifetime of the $K^{*}(892)^0$ resonance ($\sim$~4~fm/c) is comparable to the
expected duration of the rescattering hadronic gas phase between freeze-outs. Consequently, a certain fraction of $K^{*}(892)^0$ resonances will decay inside the fireball and their decay products may be significantly modified by elastic
scatterings. In such a case a suppression $K^{*}(892)^0$ production is expected. This effect was observed in
nucleus-nucleus collisions at Super Proton Synchrotron (SPS) and Relativistic Heavy Ion Collider (RHIC)
energies \cite{3, 4, 5, 6, 7, 8}. The ratio of $K^{*}/K$ production ($K^{*}$ stands for $K^{*}(892)^0$ , $\bar{K}^{*}(892)^0$
or $K^{*\pm}$  and K denotes $K^{+}$ or $K^{-}$ ) showed a decrease with increasing system size as expected due to the increasing rescattering time between chemical and kinetic freeze-out. The same effect has recently been reported also by the ALICE Collaboration at the Large Hadron Collider (LHC) energy \cite{9, 10, 11}.

The transverse mass spectra and yields of $K^{*}(892)^0$ mesons are also very important inputs for Blast-Wave models (determining the kinetic freeze-out temperature and transverse flow velocity) and Hadron Resonance Gas models (determining chemical freeze-out temperature, baryochemical potential, strangeness undersaturation factor, system volume, etc.). Those models significantly contribute to our understanding of the phase diagram of strongly interacting matter. In principle, the precise determination of transverse flow velocity is truly attractive nowadays, mainly for the reason, that recent LHC, RHIC and even SPS results suggest that a dense and collectively behaving system may appear also in collisions of small nuclei, or even in elementary interactions. Finally, the study of resonances in elementary interactions contributes to the understanding of hadron production, due to the fact 
that products of resonance decays represent a
large fraction of the final state particles. Resonance spectra and yields provide an important reference for
tuning Monte Carlo hadron production models.

In this paper we report measurements of $K^{*}(892)^0$ resonance production via its $K^{+}-\pi^{-}$ decay mode in
inelastic p+p collisions at beam momentum 158 GeV/c ( $\sqrt{s_{NN}}$ = 17.3 GeV). The data were recorded by
the NA61/SHINE hadron spectrometer \cite{21} at the CERN SPS. Unlike in the previous NA49 analysis \cite{4}
at the same beam momentum, the template fitting method was used to extract the $K^{*}(892)^0$ signal, and this method
was found to be much more effective in background subtraction than the mixing technique. 
The template analysis method is also known as the cocktail fit method and was used by many other experiments such as ALICE, ATLAS, CDF, and CMS~\cite{Aduszkiewicz:2017anm}. Moreover, the large statistics NA61/SHINE data (about 57M recorded events) allowed to obtain high-quality double differential transverse momentum and rapidity spectra of $K^{*}(892)^0$ mesons. 

\section{Methodology}
\label{s:methodology}

The $K^{*}(892)^0$ analysis was done for p+p interactions based on high-statistics data sets recorded in years 2009, 2010 and 2011) which
contained about 56.65 x $10^6$ collisions of the proton beam with a 20 cm long liquid hydrogen target. The NA61/SHINE calibration, track and vertex reconstruction procedures and simulations are discussed in Refs. \cite{22, 23, 26}

The analysis procedure is divided into the following step:

\begin{itemize}
	\item[(i)] event selection (choosing the inelastic collisions in the target with good quality fitted vertex),
	\item[(ii)] track selection (tracks from the main vertex with good momentum reconstruction and sufficient number of points in the TPCs),	 
	\item[(iii)] identification of $K^{+}$ and $\pi^{-}$ particles based on the measurement of the ionization energy loss dE/dx in the gas volume of the TPCs. The $K^{+}$ and $\pi^{-}$ candidates were selected by requiring their dE/dx values to be within 1.5$\sigma$ or 3.0$\sigma$ around their nominal Bethe-Bloch values, respectively, where $\sigma$ represents the typical standard deviation of a Gaussian dE/dx distribution of kaons and pions,
	\item[(vi)] calculation of invariant mass distributions of $K^{+}\pi^{-}$ pairs,
	\item[(v)] calculation of invariant mass distributions of $K^{+}\pi^{-}$ pairs for mixed events and Monte Carlo templates,
	\item[(vi)] extraction of the $K^{*}(892)^0$ signal using fitting procedures of the invariant mass distributions,
	\item[(vii)] calculation of correction factors for inefficiencies using MC simulated data (they include geometrical acceptance, reconstruction efficiency and corrections for losses of inelastic p+p interactions due to the trigger and the event and track selection criteria).
\end{itemize}

The raw yields of $K^{*}(892)^0$ are obtained by performing fits to the invariant mass spectra using a function described by Eq.(\ref{eq:minv}):

\begin{equation}
	f\left(m_{K^{+}\pi^{-}}\right)=a_{R} \cdot T_{resonances}^{MC}\left(m_{K^{+}\pi^{-}}\right) + b_{M}\cdot T_{mixed}^{DATA}\left(m_{K^{+}\pi^{-}}\right) + c_{S} \cdot BW \left(m_{K^{+}\pi^{-}}\right)
	\label{eq:minv}
\end{equation}

The symbols $a_{R}$, $b_{M}$ and $c_{S}$ in Eq. (\ref{eq:minv}) are the normalization parameters of the fit ($a_{R}+b_{M}+c_{S} = 1$), which describe the contributions of $T_{resonances}^{MC}$, $T_{mixed}^{DATA}$ and $BW$ to invariant mass spectra. The background is described as a sum of two elements: $T_{resonances}^{MC}$ and $T_{mixed}^{DATA}$. The $T_{mixed}^{DATA}$ template is the background description created based on the mixing method, which uses the invariant mass spectra calculated for $K^{+}$,$\pi^{-}$ pairs from different events. The $T_{resonances}^{MC}$ is the background, which describes the contribution of $K^{+}\pi^{-}$ pairs coming from other resonances with exception of $K^{*}(892)^0$. It is defined as the sum of:

\begin{itemize}
	\item combination of tracks that come from decays of various resonances, for example one track from a $\rho^0$ meson and one from a $K^{*+}$ meson,
	\item combination of tracks where one comes from decay of a resonance and one comes from a direct
	production in the interaction vertex.
\end{itemize}

The $T_{resonances}^{MC}$ templates were constructed by passing Monte Carlo simulated p+p production interactions, generated with
the  EPOS~1.99~\cite{27} hadronic interaction model using the CRMC 1.4 package \cite{28}, through the full NA61/SHINE detector Monte Carlo chain and then through the same reconstruction routines as the data. For the
reconstructed simulated data, the same event and track selection criteria, as for real data, were used. Both the templates and the data were analyzed in rapidity ($y$) intervals (calculated in the center-of-mass reference system) and transverse momentum $p_T$.

Finally, the signal (BW) is described using the Breit-Wigner distribution:

\begin{equation}
	BW \left(m_{K^{+}\pi^{-}}\right) =A \cdot \frac{\frac{1}{4} \cdot \Gamma_{K^*}^2}{(m_{K^{+}\pi^{-}} - m_{K^*})^2 + \frac{1}{4} \Gamma_{K^{*}}^2}
\end{equation}

where A is the normalization factor. The initial values for the parameters of the mass ($ m_{K^*}$) and width
($\Gamma_{K^{*}}$) of $K^* (892)^0$ were taken as the PDG values: $ m_{K^*}$ = 0.89555 GeV and $\Gamma_{K^{*}}$ = 0.0473 GeV \cite{29}. 

\begin{figure}
	\includegraphics[width=0.5\textwidth]{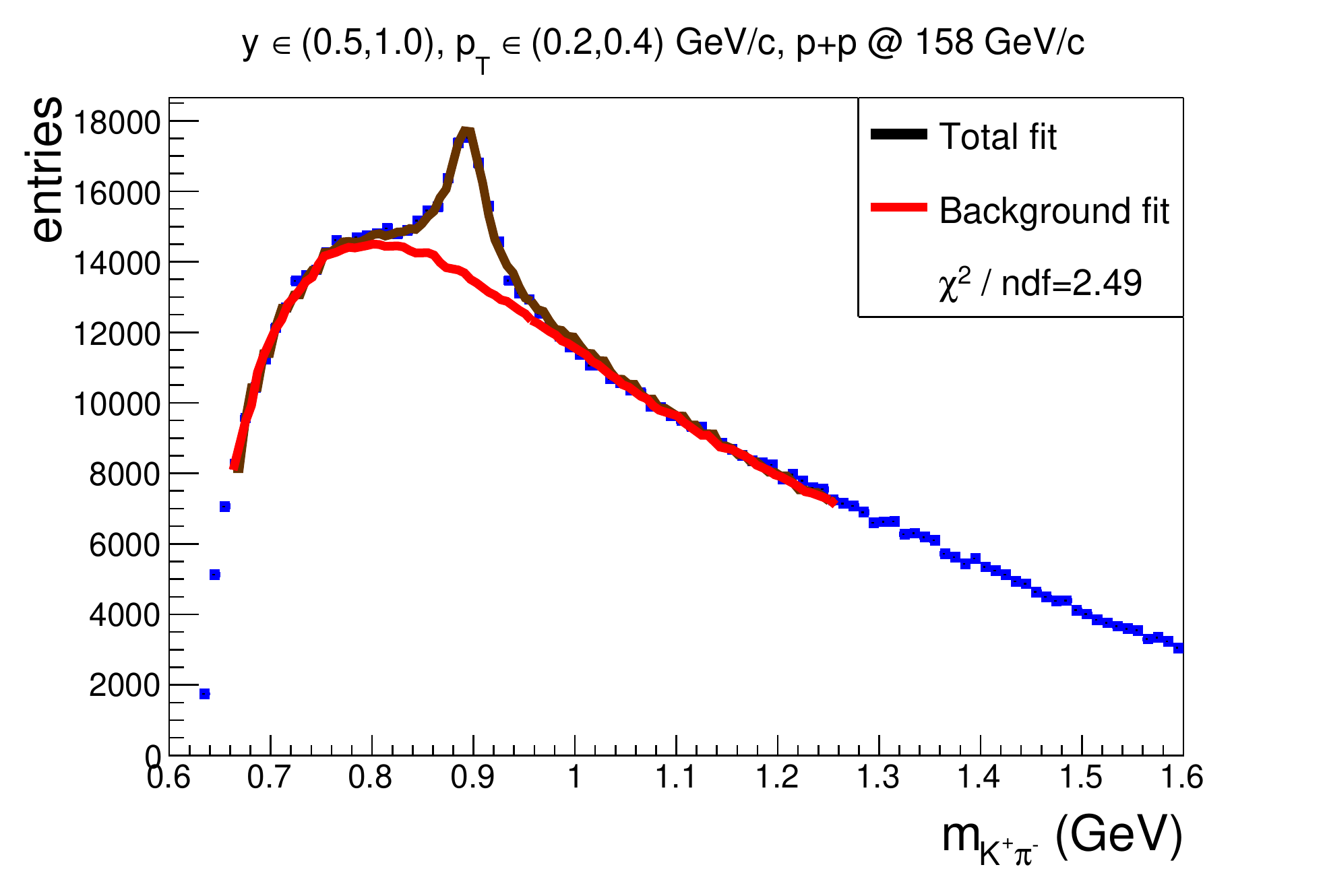} 
	\includegraphics[width=0.5\textwidth]{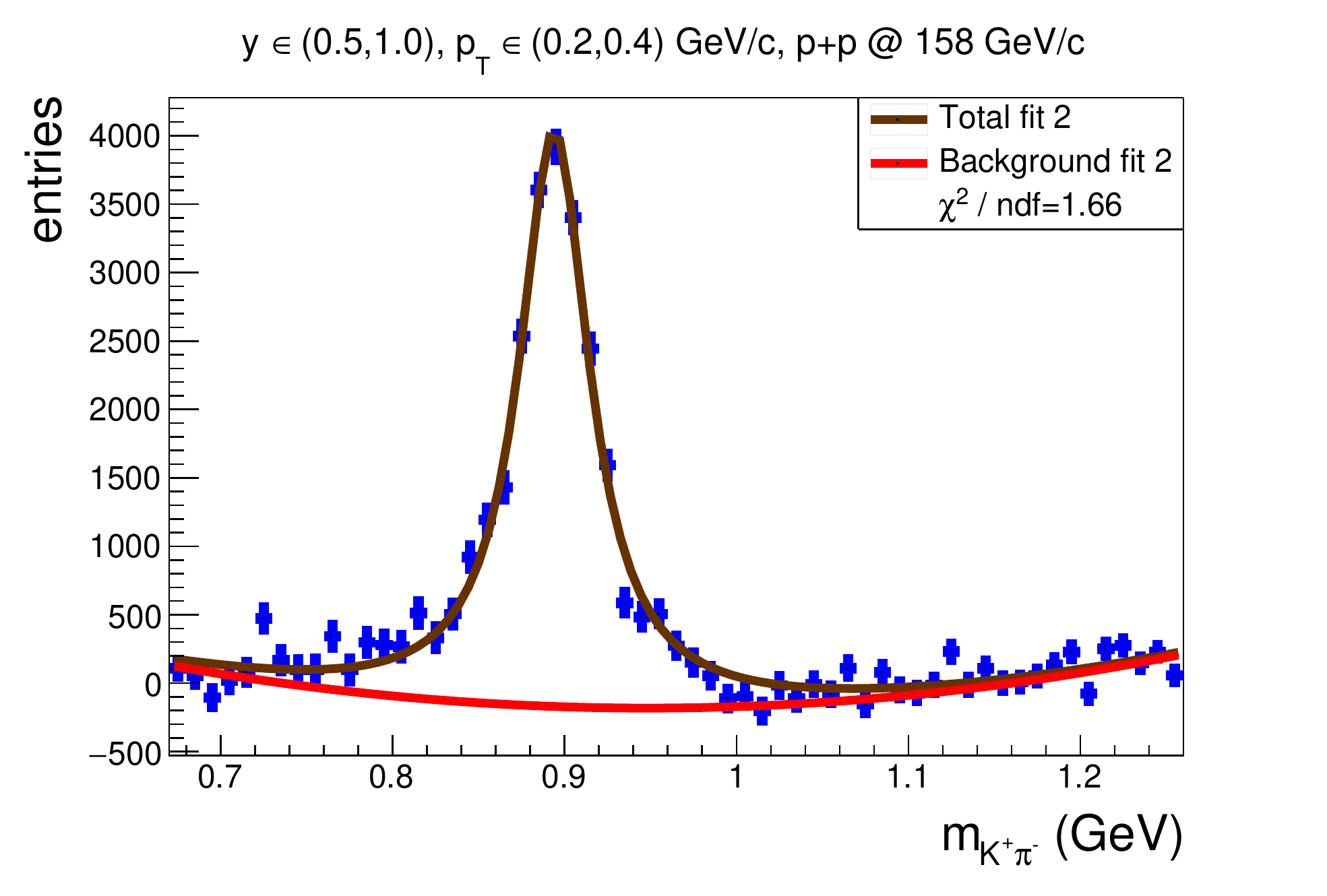}
	\caption{(Color online) Example of the procedure of signal extraction for $K^*(892)^0$ in rapidity bin 0.5 < y <
		1.0 (all rapidity values in the paper are given in the center-of-mass reference system) and transverse momentum
		bin 0.2 < $p_T$ < 0.4 GeV/c for p+p collisions at 158 GeV/c. Left: data (blue points) and estimated background (red
		histogram) obtained from the templates. Right: background subtracted signal for template method.}
		\label{fig:minv}
\end{figure}

In Fig. \ref{fig:minv} (left), the fitted invariant mass spectrum, using Eq. (\ref{eq:minv}), is presented by a brown curve.
The red line shows the fitted function but without the signal description element (BW). Both fits were
performed in the invariant mass range from 0.66 GeV to 1.26 GeV. After subtraction of Eq. (\ref{eq:minv}) without BW element (red curve in Fig. \ref{fig:minv} (left)), the mass distribution is shown in Fig. \ref{fig:minv} (right). The red line here is an additional fitted background contribution parameterized by a second order polynomial curve. In fact, a rudimentary background is present only for the $y$ and $p_T$ bins in which the statistics is very low (not shown here). After subtraction of the fitted second order polynomial curve
the resulting signal distribution (done for all $y$ and $p_T$ bins), is fitted with the Breit-Wigner shape in
the mass window $m_{K^*} \pm 4 \Gamma_{K^*}$ to obtain the final values of mass and width of $K^*(892)^0$. The uncorrected
numbers of $K^*(892)^0$ are obtained by integrating the signal distribution in the mass window $m_{K^*} \pm 4 \Gamma_{K^*}$.

\begin{figure}
		\centering
		\includegraphics[width=0.49\textwidth]{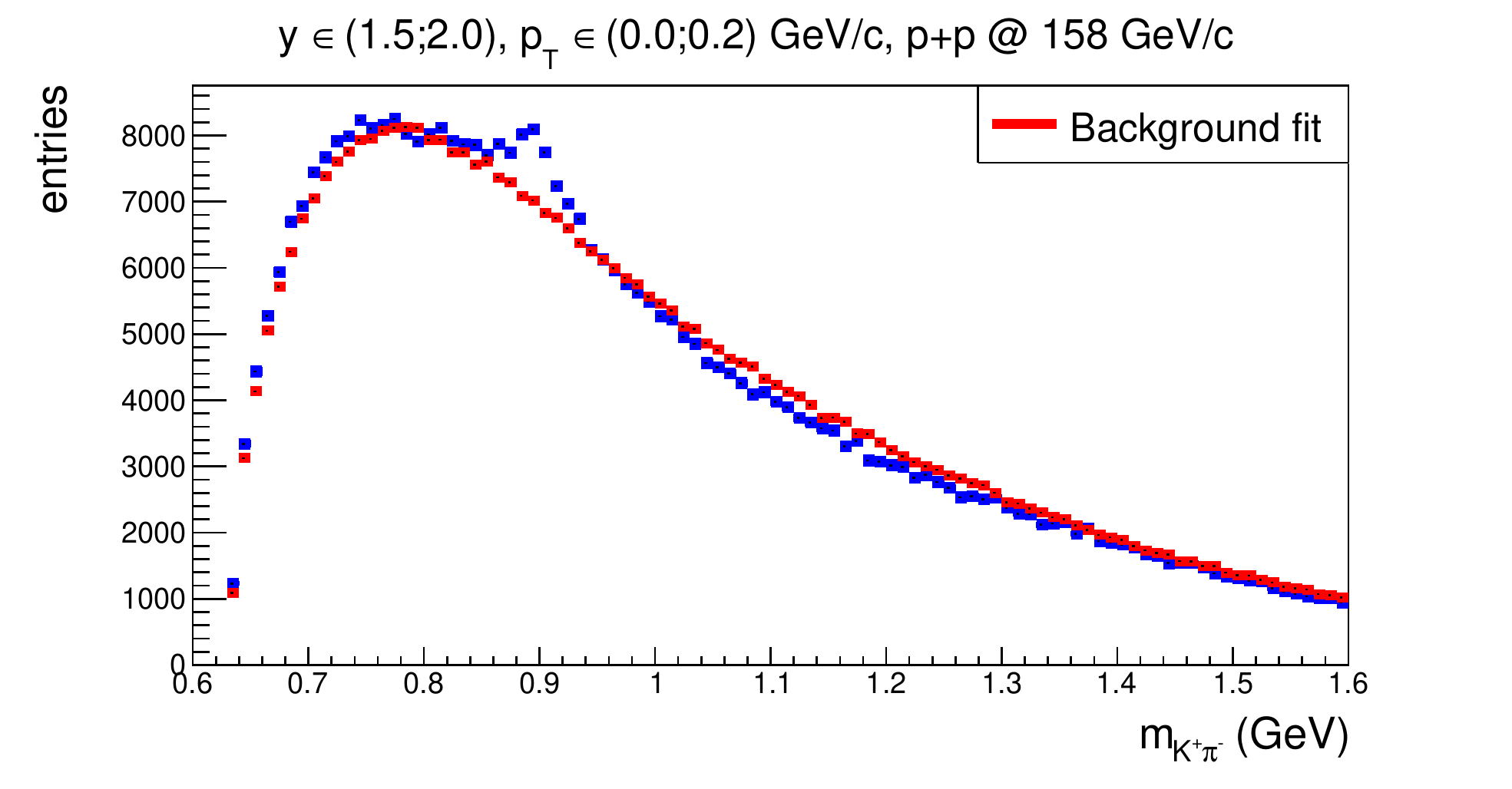} 
		\includegraphics[width=0.49\textwidth]{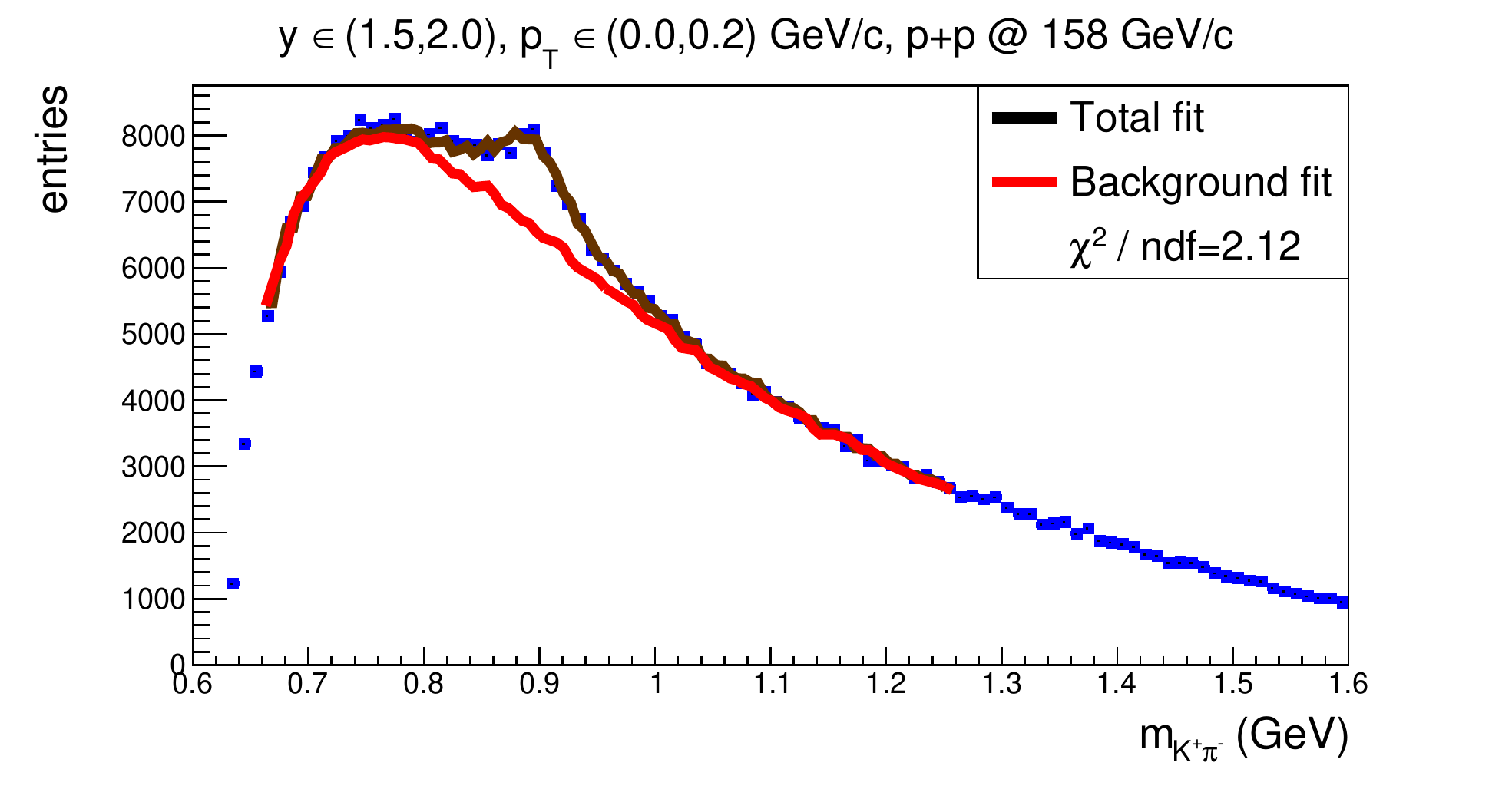} \\
		\includegraphics[width=0.49\textwidth]{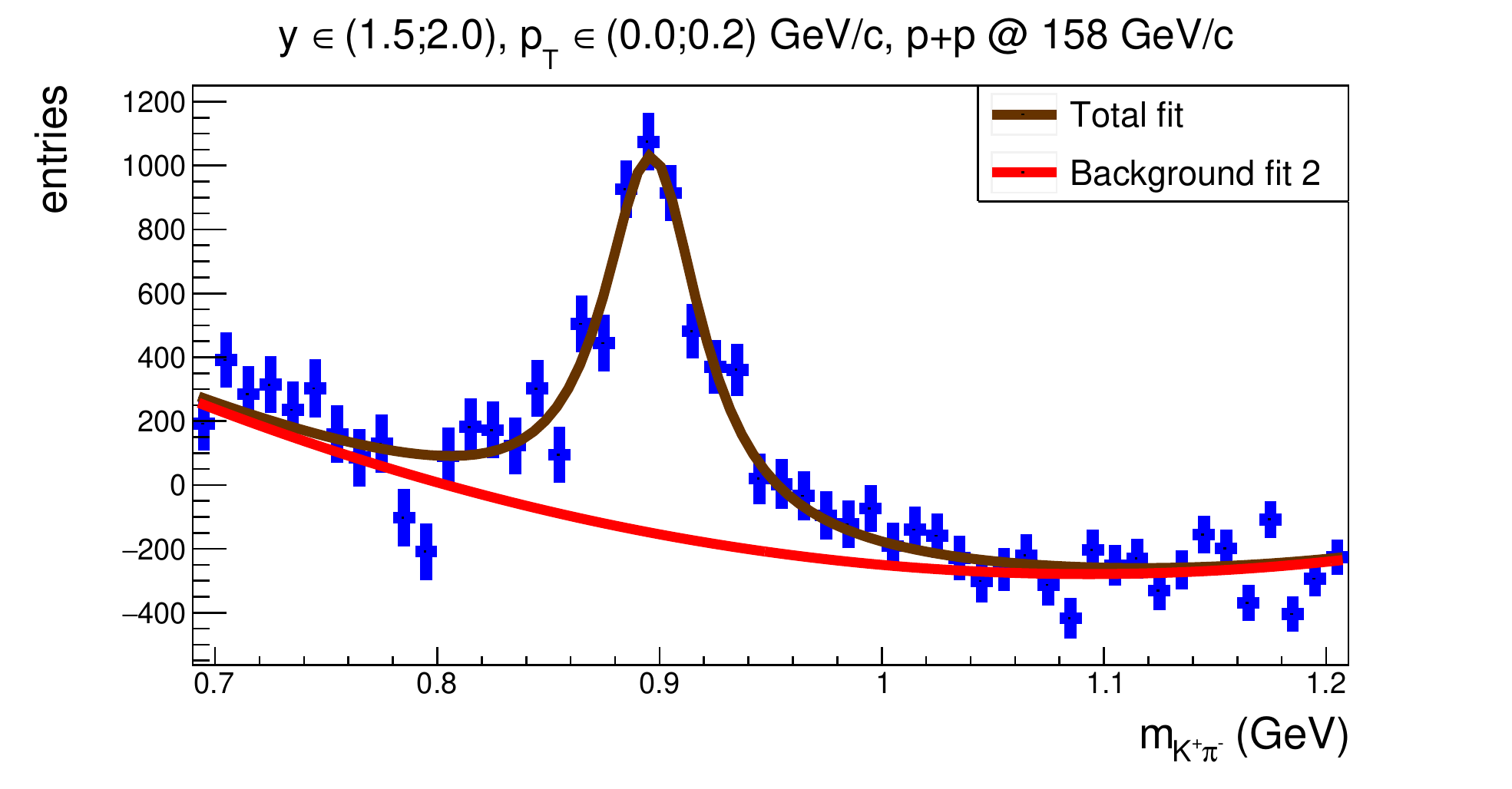} 
		\includegraphics[width=0.49\textwidth]{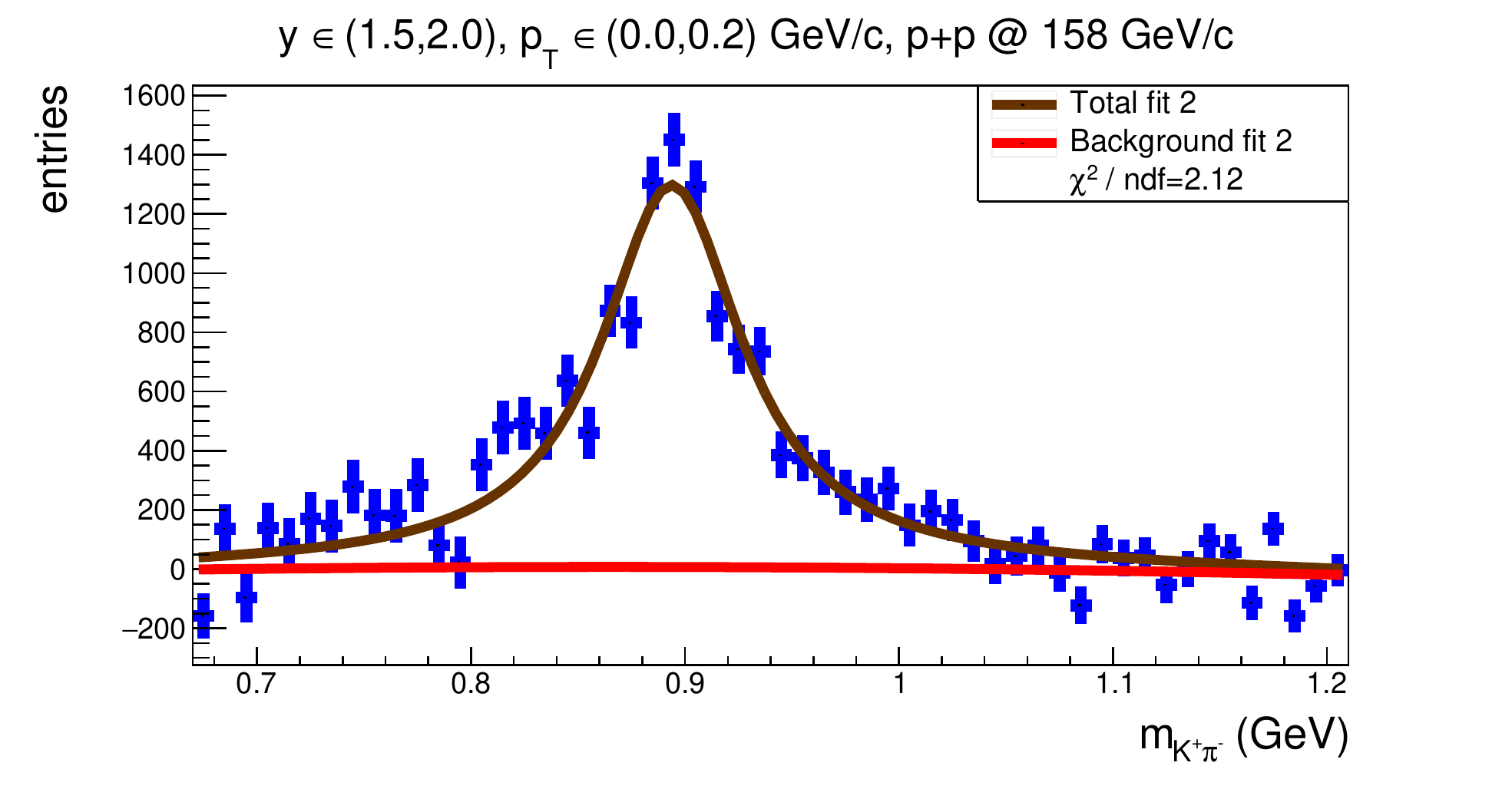} \\
		\label{fig:comparison}
		\caption{(Color online) Example of the procedure of signal extraction for $K^* (892)^0$ in rapidity bin 1.5 < $y$ <
			2.0 and transverse momentum bin 0.0 < $p_T$ < 0.2 GeV/c for p+p collisions at 158 GeV/c. Top, left: data (blue points) and fitted background (red histogram) obtained from mixed events (standard method). Top, right: data (blue points) and background (red line) obtained from the templates. Bottom: background subtracted signal for the standard method (left) and template method (right).}
\end{figure}

The comparison between the mixing technique of describing the background (based only on mixed events) to the template analysis method is shown in Fig \ref{fig:comparison}. 

In order to determine the number of $K^*(892)^0$ produced in inelastic p+p interactions, two corrections were applied to the extracted raw number of $K^* (892)^0$:
\begin{itemize}
	\item[(i)] The loss of the $K^* (892)^0$ due to the dE/dx requirement, was corrected by a constant factor: 
	
	\begin{equation}
	c_{dE/dx} = \frac{1}{\epsilon_{K^+}\epsilon_{\pi^-}}
	\end{equation}
	
	where $\epsilon_{K^+}$ = 0.866, $\epsilon_{\pi^-}$ = 0.997 are the probabilities for $K^+$ or $\pi^-$ to lie within 1.5$\sigma$ or 3$\sigma$ around the nominal Bethe-Bloch value.
	
	\item[(ii)] A detailed Monte Carlo simulation, based on about 83.8 x $10^6$ collisions,	was performed to correct for geometrical acceptance, reconstruction efficiency, losses due to the trigger bias, detector acceptance as well as the quality cuts applied in the analysis (correction factor is defined in the Eq.~\ref{eq:cmc}).

	\begin{equation}
	c_{MC}(y,p_{T}) = \frac{n_{gen}}{n_{sel}}
	\label{eq:cmc}
	\end{equation}
where $n_{gen}$ and $n_{sel}$ are numbers of simulated and reconstructed $K^{*}(892)^0$ per inelastic event.

\end{itemize}
\section{Double differential spectra}

The double differential yield of  $K^* (892)^0$ per inelastic event in a bin of $(y,p_T)$ is calculated as follows:

\begin{equation}
	\frac{d^2 n}{dy dp_T} (y, p_T) = \frac{1}{BR} \cdot \frac{N_{K^*}(y,p_T)}{N_{events}} \cdot \frac{c_{dE/dx}\cdot c_{MC}(y,p_T)}{\Delta y \Delta p_T}
\end{equation}

where:
\begin{itemize}
\item[-] BR = 2/3 is the branching ratio of $K^* (892)^0$ decay into $K^{+}\pi^-$
\item[-] $N_K^∗(y, p_T )$ is the uncorrected number of $K^* (892)^0$, obtained by the signal extraction procedure described in Section \ref{s:methodology},
\item[-] $N_{events}$ is the number of events after cuts,
\item[-] $c_{dE/dx}$ , $c_{MC} (y, p_T )$ are correction factor described above,
\item[-] $\Delta y$ and $\Delta p_T$ are the bin widths.
\end{itemize}

\begin{figure}
	\centering
	\includegraphics[width=\textwidth]{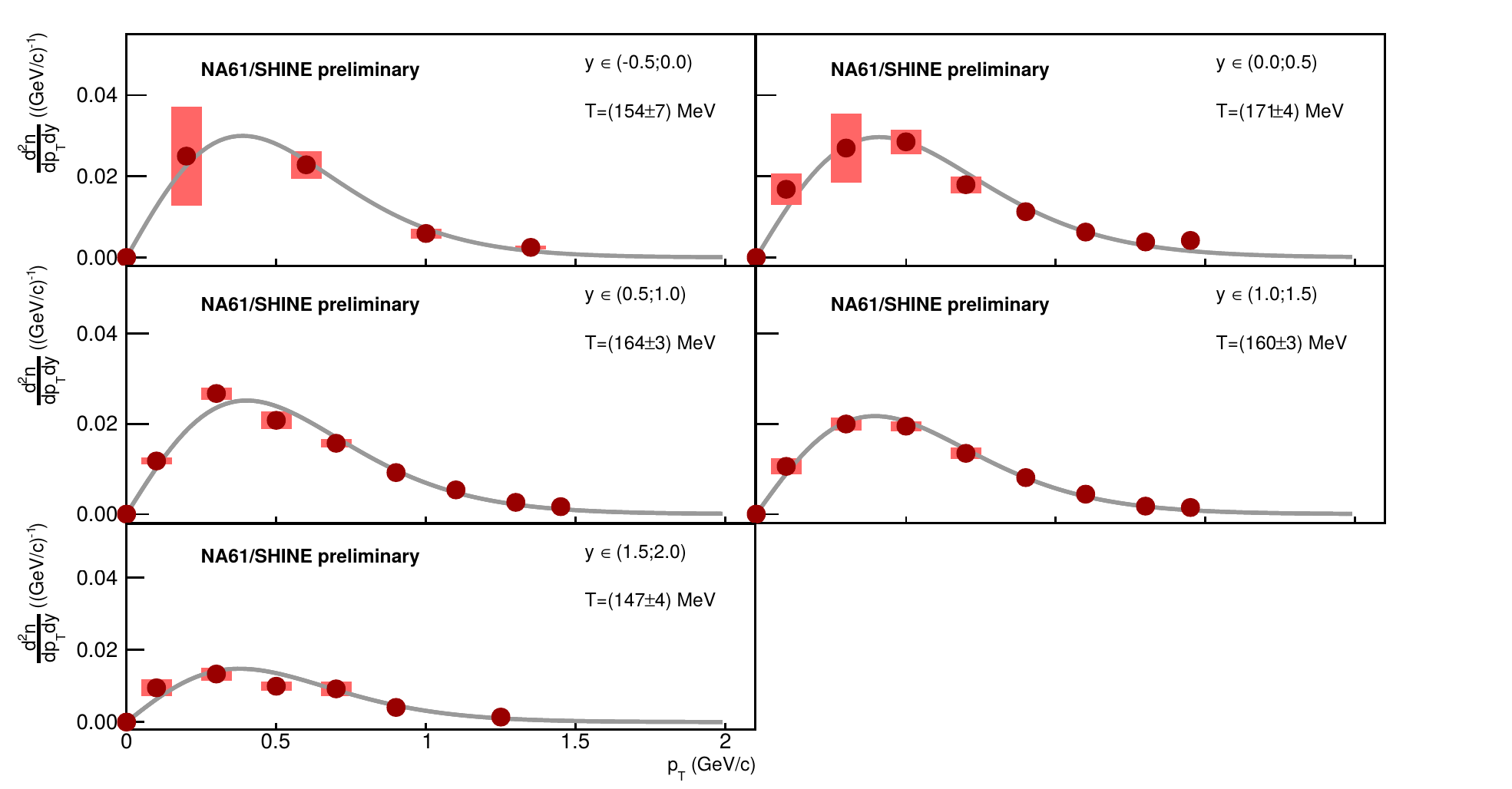}
	\label{fig:dndydpt}
	\caption{(Color online) Transverse momentum spectra $\frac{d^2 n}{dy dp_T}$ for five bins of rapidity. The fitted function is given by Eq. (\ref{eq:fit}). The fitted inverse slope parameters for each bin are in the legends.}
\end{figure}

Figure \ref{fig:dndydpt} shows the double differential yields of $K^* (892)^0$ mesons presented for separate rapidity bins. In order to measure the inverse slope parameters ($T$) of transverse momentum spectra and, later on, to estimate the yield of $K^* (892)^0$ mesons in the unmeasured high $p_T$ region, the function given by Eq. (\ref{eq:fit}) was fitted to the data from Fig. (\ref{fig:dndydpt}). The quantity $m_T ≡ \sqrt{p_T^2 + m^2_{K^∗}}$ represents the transverse mass of $K^* (892)^0$, where $m_{K^*}$ is the PDG value.

\begin{equation}
	f(p_T) = A \cdot p_T exp \left ( - \frac{\sqrt{p^2_T + m^2_{K^{*}}}}{T} \right)
	\label{eq:fit}
\end{equation}

\section{Rapidity spectrum}

The rapidity distribution $\frac{dn}{dy}$ was calculated by integrating the measured and extrapolating the non-measured $p_T$ region of the $\frac{d^2 n}{dy dp_T}$ spectrum according to Eq. (\ref{eq:dndy}).

\begin{equation}
	\frac{dn}{dy} = \left ( 1 + \frac{A_{p_T}}{I_{p_T}} \right ) \sum_{i} \frac{d^2 n}{dy dp_T} \cdot p_T
	\label{eq:dndy}
\end{equation}

where: \\

\begin{equation}
A_{p_T} = \int_{1.5}^{+\infty} A \cdot p_T exp \left ( - \frac{\sqrt{p^2_T + m^2_{K^{*}}}}{T} \right) dp_T, \hspace{0.7cm} I_{p_T} = \int_{0}^{1.5} A \cdot p_T exp \left ( - \frac{\sqrt{p^2_T + m^2_{K^{*}}}}{T} \right) dp_T,
\end{equation}

\begin{figure}
	\includegraphics[width=\textwidth]{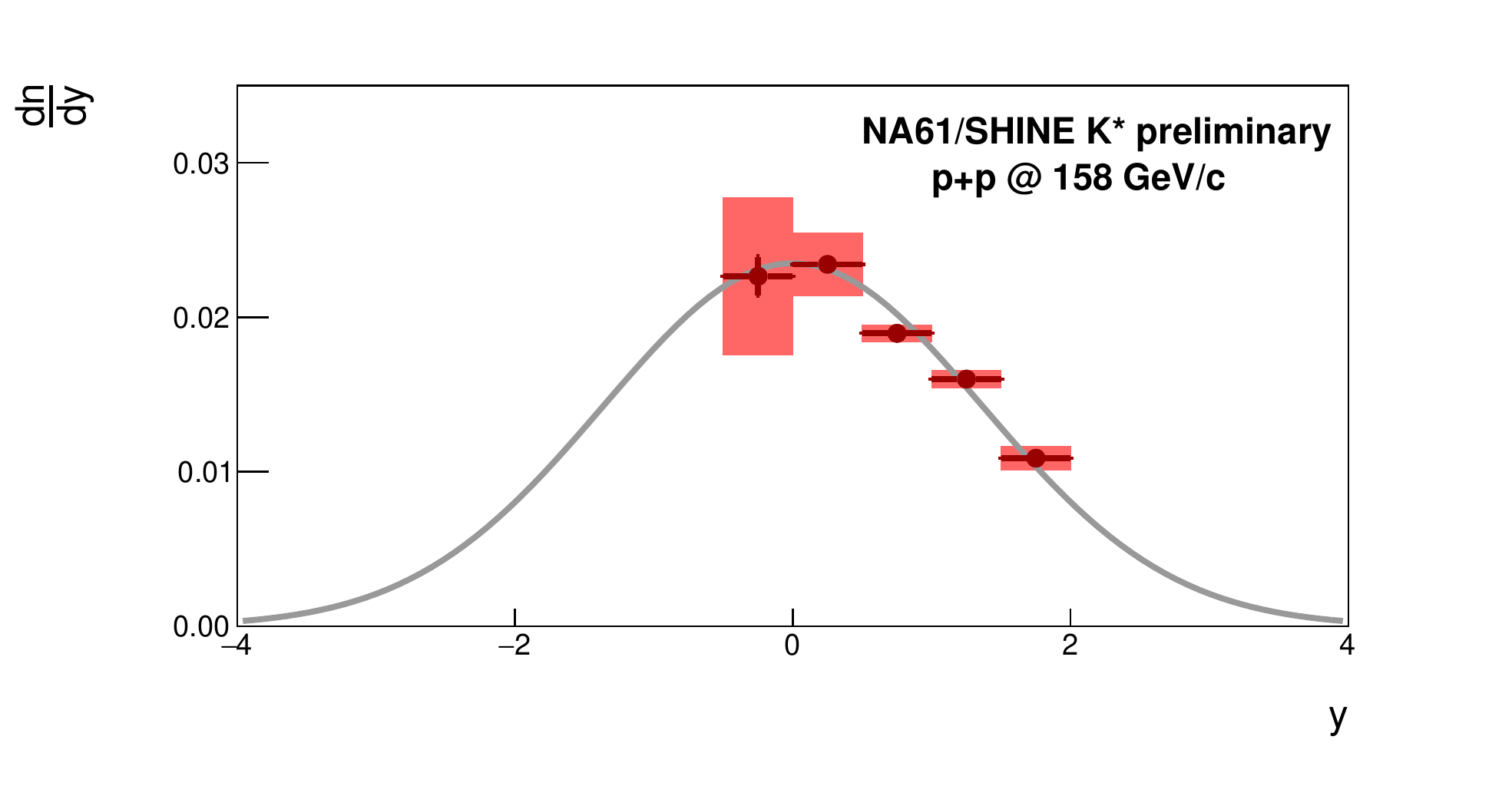}
	\label{fig:dndy}
\caption{(Color online) The $p_T$-integrated and extrapolated rapidity distribution. The fitted Gaussian function is given by Eq.
(\ref{eq:fit2}). The NA61/SHINE numerical data are listed in the Table \ref{tab:results}.
}
\end{figure}

The $p_T$-integrated and extrapolated $\frac{dn}{dy}$ spectrum of $K^* (892)^0$ mesons is presented in Fig. \ref{fig:dndy}. The Gaussian function given by Eq. (\ref{eq:fit2}) was fitted to the data to measure the width $\sigma_y$ and total yield of $K^* (892)^0$ in inelastic p+p collisions at 158 GeV/c. The numerical values are shown in Table \ref{tab:results}. The NA49 results are taken from \cite{4}.

\begin{equation}
	f(y) = A \cdot exp \left ( - \frac{y^2}{2 \sigma_y^2} \right)
	\label{eq:fit2}
\end{equation}

\begin{table}[h]
	\centering
	\begin{tabular}{|c|c|c|}
		\hline
		& NA61 & NA49 \\
		\hline
		$\langle K^{*}(892)^o \rangle$ & 0.08058 $\pm$ 0.00059 $\pm$ 0.00260  & 0.0741 $\pm$ 0.0015 $\pm$ 0.0067 \\
		\hline
	\end{tabular}
	\caption{The mean multiplicity of $K^* (892)^0$ was calculated as the integral under the fitted function 
Eq.~\ref{eq:fit2}, where the range of integration in NA61/SHINE was selected as -4 < y < 4.}
	\label{tab:results}
\end{table}

\section{Mass and width of $K^{*}(892)^0$}

\begin{figure}
	\centering
	\includegraphics[width=0.9\textwidth]{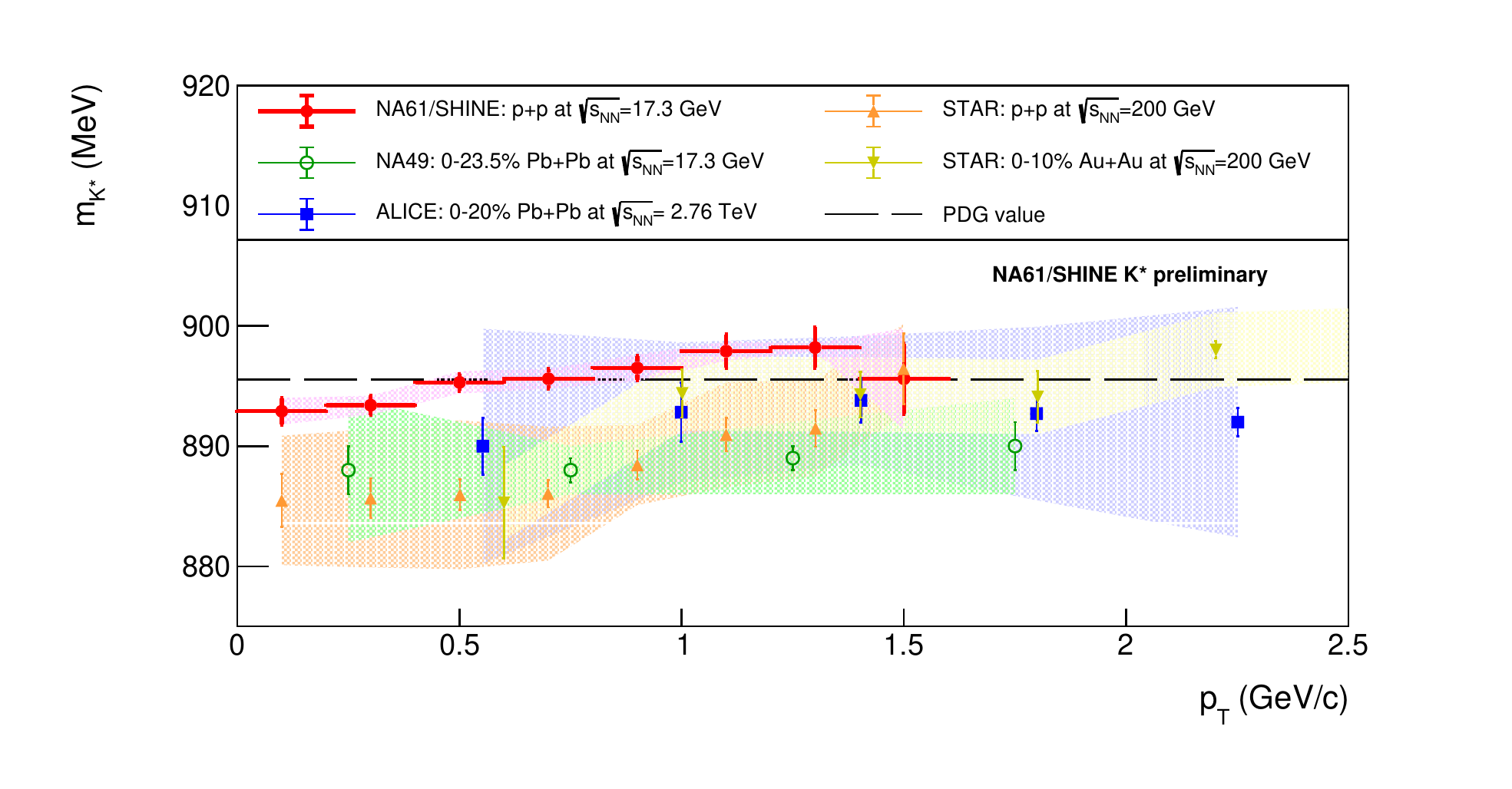} \\
	\includegraphics[width=0.9\textwidth]{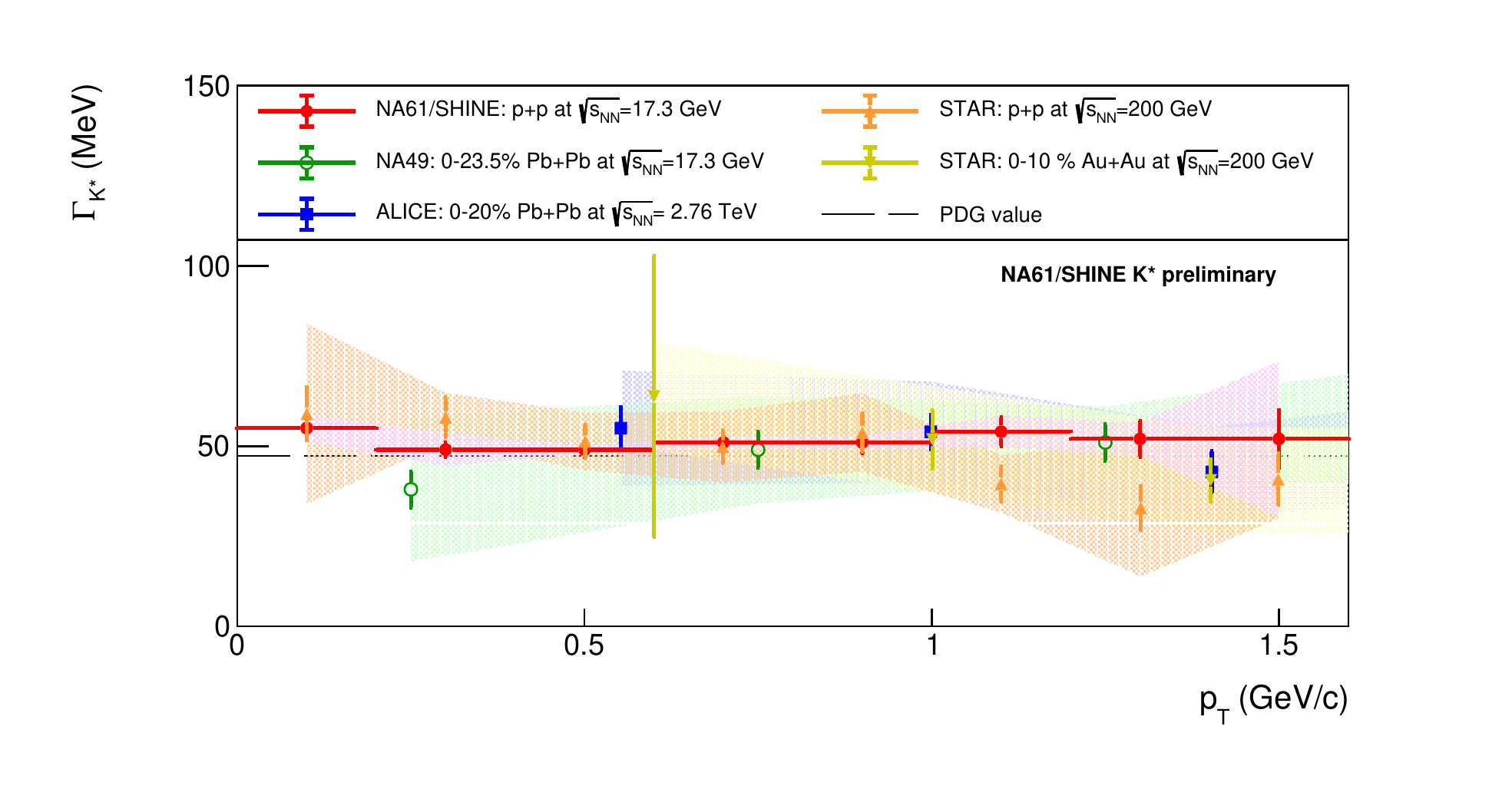} 
	\label{fig:mass}
	\caption{(Color online) The transverse momentum dependence of mass and width of $K^* (892)^0$ mesons from NA61/SHINE, NA49 \cite{4}, ALICE \cite{9} and STAR \cite{7}. For ALICE and STAR the averaged ($K^{*0}$ ) measurements of $K^* (892)^0$ and $\bar{K^*} (892)^0$ are shown. The horizontal lines represent PDG values \cite{29}.}
\end{figure}

Figure \ref{fig:mass} shows the comparison of mass and width of $K^* (892)^0$ mesons obtained in NA61/SHINE p+p
collisions, STAR p+p data (top RHIC energy), as well as in Pb+Pb and Au+Au interaction at SPS, RHIC
and LHC energies. For ALICE and STAR experiments the averaged measurements of $K^* (892)^0$ and
$\bar{K^*} (892)^0$ mesons are shown. One sees that among the presented results (and within the $p_T$ range shown
in the figure) the precision of the NA61/SHINE measurements is the highest and the results are very close
to the PDG values. For p+p collisions the STAR experiment measured lower $K^{*0}$ mass, especially at lower transverse momentum region.

\section{Comparisons}

The statistical Hadron Resonance Gas Models (HGM) are commonly used to predict particle multiplicities in elementary and nucleus-nucleus collisions, using as adjustable parameters the chemical freeze-out temperature $T_{chem}$,
the baryochemical potential $\mu_B$ , strangeness saturation parameter $\gamma_S$ , etc. In this paper the $\langle K^∗ (892)^0 \rangle$ values are compared with predictions of two HGM models described in Refs. \cite{32,33}.

\begin{figure}
	\centering
	\includegraphics[width=0.6\textwidth]{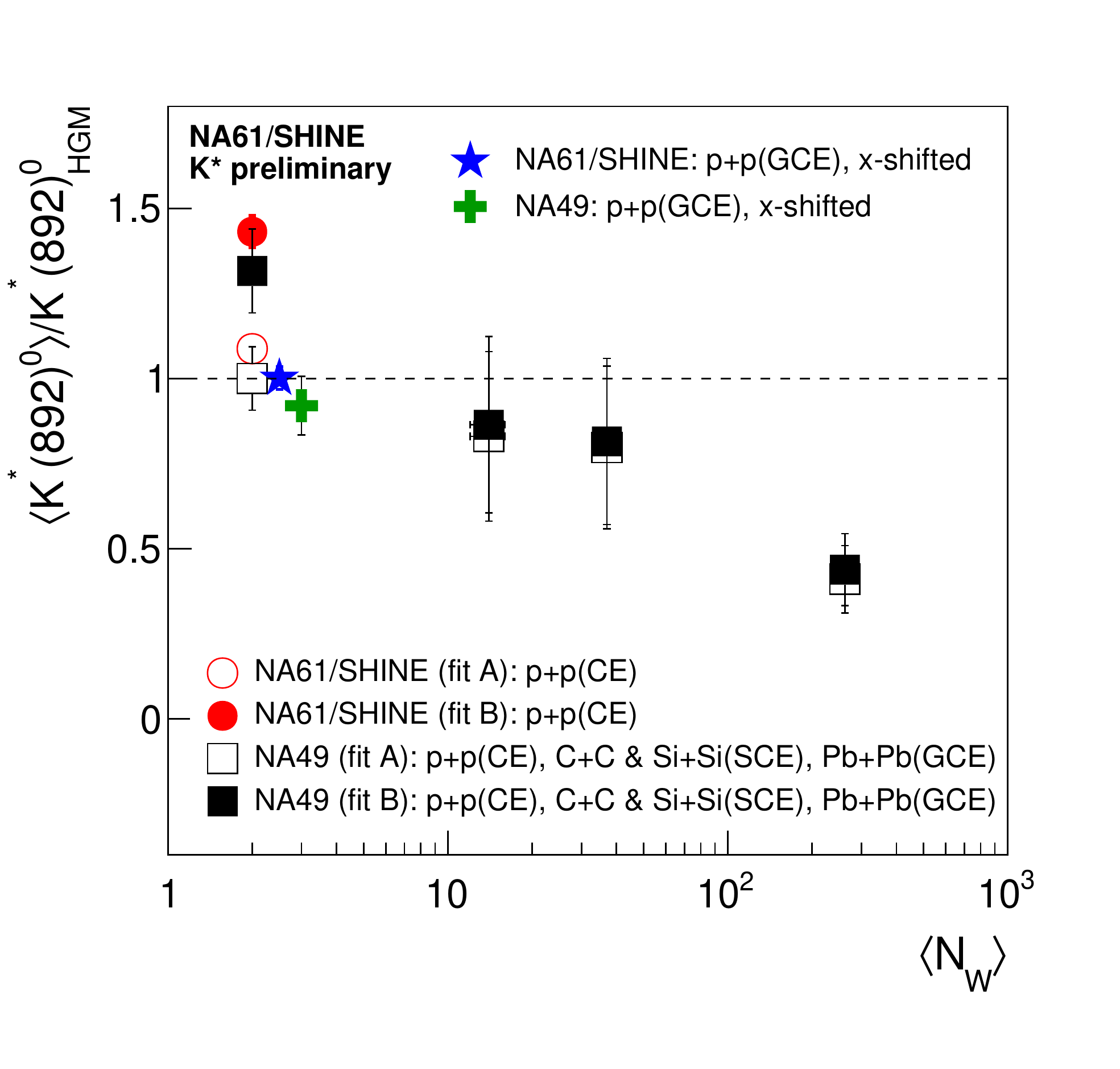}
	\label{fig:HGM}
	\caption{(Color online) The mean multiplicity of $K^∗ (892)^0$ for p+p (this analysis and NA49 \cite{4}), as well as NA49
		C+C, Si+Si and Pb+Pb \cite{4} for 158A GeV/c collisions divided by the HGM predictions for fit B, fit A and Grand Canonical Ensemble formulation. $N_W$ denotes the number of wounded nucleons and its values are taken from Ref. \cite{4}. HGM predictions are taken from Ref. \cite{32,33,34}.}
\end{figure}

In Ref. \cite{32} the HGM model predictions were performed for two versions of the fits. The first one called
fit B, allowed strangeness under-saturation so the usual parametrization with $\gamma_S$ was
applied. For p+p data, the fit was obtained with removing the $\Xi^{'}_S$ and $\Omega^{,}_S$ baryons from the data sample. In the second fit, called A, the parameter $\gamma_S$ was replaced by the mean number of strange quark pairs $\langle s \bar{s} \rangle$. For p+p data, fit A was obtained with removing the $\phi$ meson from the data sample. For both fits, in case of p+p collisions, theoretical multiplicities were calculated in the Canonical Ensemble (CE) \cite{32}. The mean multiplicity of $K^* (892)^0$ for 158 GeV/c inelastic p+p interactions was divided by HGM predictions based on fit A and B, separately, and compared with the value for NA49 data \cite{4}. The results are shown in Fig \ref{fig:HGM} for p+p data, as well as C+C, Si+Si, and Pb+Pb interactions measured by NA49 \cite{4}. In Ref. \cite{32} for heavier C+C and Si+Si systems the S-Canonical Ensemble was used (assumes exact strangeness
conservation and grand-canonical treatment of electric charge and baryon number), and for Pb+Pb the
Grand Canonical Ensemble (GCE) was assumed. For C+C and Si+Si interactions, all available particles
were used in the HGM fits, including $\phi$ meson and multi-strange baryons. For Pb+Pb data only the
measured $\Lambda$(1520) yield was removed from the fitted data sample. Please note that the centrality of
Pb+Pb collisions used in the HGM fits was 0-5\% whereas the $\langle K^∗ (892)^0 \rangle$ values in NA49 were obtained
for the 0-23.5\% most central interactions. Therefore, the HGM yields had to rescaled by a factor 262/362, before comparing them to the NA49 data.

For heavier systems (including C+C and Si+Si), there is no significant difference between fit A and fit B, however,
the deviation between the HGM predictions and experimental data increases with increasing system size. The p+p data are very close to HGM prediction but only in case of fit A, where the $\phi$ meson was excluded from the fit. In the most recent paper \cite{33}, where the HGM fits were done for old NA49 and new NA61/SHINE p+p data, it is also stressed that at SPS energies the $\phi$ meson multiplicities in p+p collisions cannot be well fitted within the CE formulation of the HGM (quality of CE fits becomes much worse when the $\phi$ meson yield is included). But the mean multiplicity of $K^∗ (892)^0$ mesons in inelastic p+p collisions at 158 GeV/c can be compared to the HGM prediction based on the Grand Canonical Ensemble formulation \cite{33}. The results for NA49 and NA61/SHINE data are shown in Fig. \ref{fig:HGM}. Surprisingly, the GCE statistical model provides a very good description of the $K^∗ (892)^0$ yield in the small p+p system. 

The $K^*$ to charged kaons ratios, may allow studying the length of the time interval between chemical and kinetic freeze-out in
nucleus-nucleus collisions. The $K^*$ and K mesons have identical quark (anti-quark) contents, but
different mass and relative orientation of quark spins. Thus, the $ \langle K^∗ (892)^0 \rangle/\langle K^− \rangle$ and $ \langle K^∗ (892)^0 \rangle/\langle K^+ \rangle$ ratios are considered the least model dependent ratios for studying the $K^*$ production properties as well as the freeze-out conditions.

The NA61/SHINE $ \langle K^∗ (892)^0 \rangle/\langle K^+ \rangle$ and $\langle K^∗ (892)^0 \rangle/\langle K^− \rangle$ yield ratios for p+p collisions can be used to estimate the time between chemical and kinetic freeze-out in Pb+Pb reactions. Following Ref. \cite{7}:

\begin{equation}
\frac{K^*}{K} |_{kinetic} = \frac{K^*}{K} |_{chemical} \cdot e^{-\frac{\Delta t}{\tau}}
\end{equation}

where: \\
\begin{itemize}
	\item $ \langle K^∗ (892)^0 \rangle/\langle K^\pm \rangle$ in inelastic p+p interactions can be treated as the one at chemical freeze-out,
	\item $ \langle K^∗ (892)^0 \rangle/\langle K^\pm \rangle$ for central Pb+Pb (NA49) interactions can be used as the one at kinetic freeze-out,
	\item  $\tau$ is the $K^∗ (892)^0 $ lifetime of 4.17 fm/c \cite{29},
	\item $\Delta t$ is the time between chemical and kinetic freeze-outs.
\end{itemize}

Assuming that the losses of $K^∗ (892)^0 $ before kinetic freeze-out are due to rescattering effects and there are
no regeneration processes, the time between chemical and kinetic freeze-out can be estimated as $3.8 \pm 1.1$
fm/c from the $ \langle K^∗ (892)^0 \rangle/\langle K^+ \rangle$ ratio and $3.3 \pm 1.2$ fm/c from the  $\langle K^∗ (892)^0 \rangle/\langle K^− \rangle$ ratio. These numbers correspond to 23.5\% of the most central Pb+Pb interactions but the time would be even larger for 5\% of the most central events. The value of $\Delta t$ is larger at SPS than $\Delta t = 2 \pm 1$ fm/c obtained by RHIC \cite{7}, suggesting that regeneration effects may start to play a significant role for higher energies. As the $K^∗ (892)^0 $ regeneration may happen also at SPS energies, the obtained $\Delta t$ values can be treated rather as the lower limit of the time between chemical and kinetic freeze-out.

\begin{figure}
	\centering
	\includegraphics[width=0.7\textwidth]{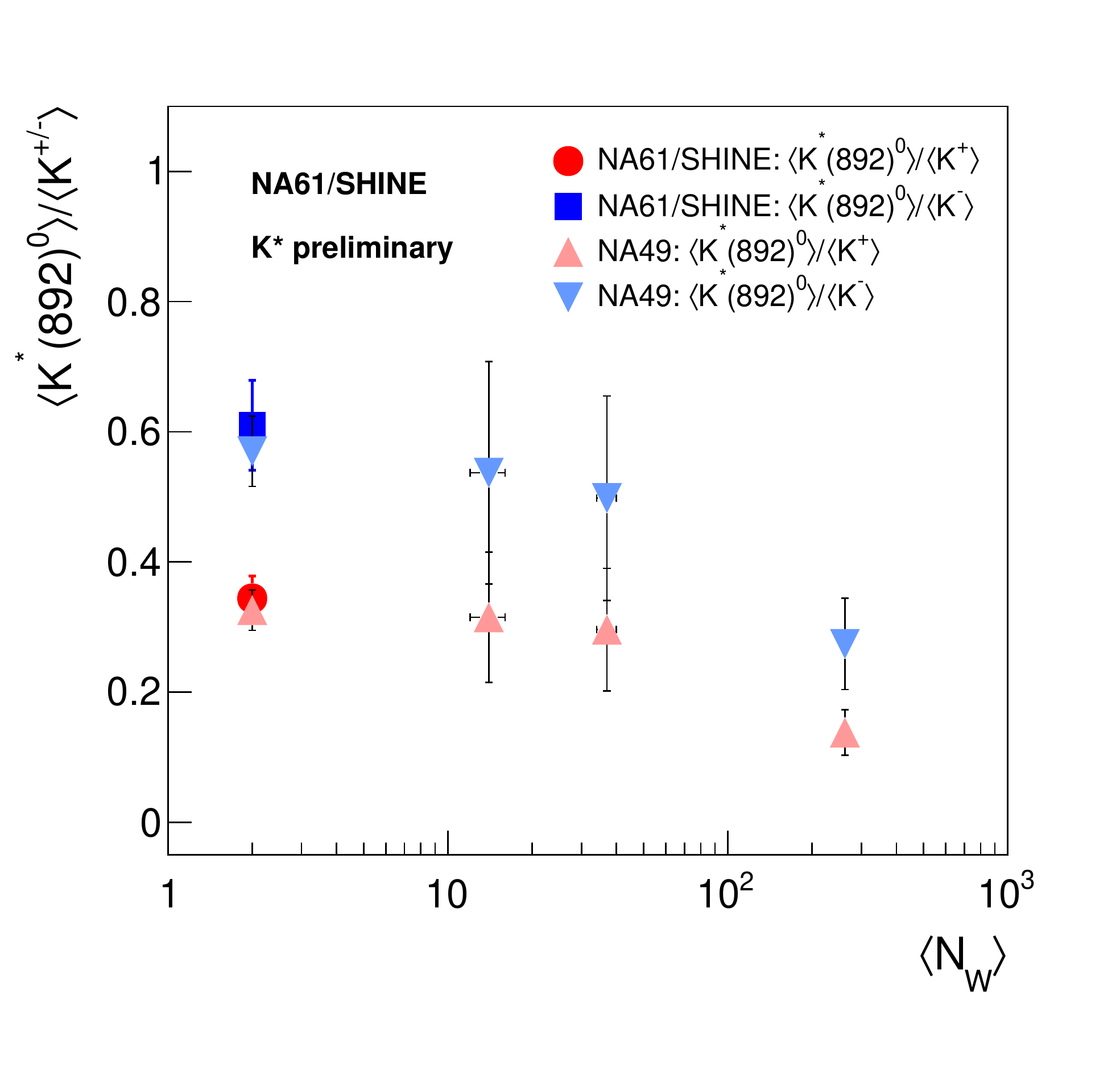}
	\caption{(Color online) The system size dependences of the $ \langle K^∗ (892)^0 \rangle/\langle K^+ \rangle$ and  $ \langle K^∗ (892)^0 \rangle/\langle K^− \rangle$ yield ratios in p+p, C+C, Si+Si and Pb+Pb collisions at 158A GeV. $N_W$ denotes the number of wounded nucleons and its values are taken from Ref. \cite{4}.}
\end{figure}

\end{document}